\documentclass[superscriptaddress,prd,aps,preprint]{revtex4}
\usepackage{graphicx,amsfonts,amsmath}
\usepackage{dcolumn}
\usepackage{bm}

\begin{document}
\title{On the finiteness of the noncommutative supersymmetric Maxwell-Chern-Simons theory}
\author{A. F. Ferrari}
\author{M. Gomes}
\affiliation{Instituto de F\'{\i}sica, Universidade de S\~ao Paulo\\
Caixa Postal 66318, 05315-970, S\~ao Paulo, SP, Brazil}
\email{mgomes,tmariz,ajsilva@fma.if.usp.br}
\author{J. R. Nascimento} 
\affiliation{Instituto de F\'{\i}sica, Universidade de S\~ao Paulo\\
Caixa Postal 66318, 05315-970, S\~ao Paulo, SP, Brazil}
\affiliation{Departamento de F\'{\i}sica, Universidade Federal da Para\'{\i}ba\\
 Caixa Postal 5008, 58051-970, Jo\~ao Pessoa, Para\'{\i}ba, Brazil}
\email{jroberto,petrov,edilberto@fisica.ufpb.br}
\author{A. Yu. Petrov}
\affiliation{Departamento de F\'{\i}sica, Universidade Federal da Para\'{\i}ba\\
 Caixa Postal 5008, 58051-970, Jo\~ao Pessoa, Para\'{\i}ba, Brazil}
\email{jroberto,passos,petrov@fisica.ufpb.br}
\author{A. J. da Silva}
\affiliation{Instituto de F\'{\i}sica, Universidade de S\~ao Paulo\\
Caixa Postal 66318, 05315-970, S\~ao Paulo, SP, Brazil}
\email{mgomes,ajsilva@fma.if.usp.br}
\author{E. O. Silva}
\affiliation{Departamento de F\'{\i}sica, Universidade Federal da Para\'{\i}ba\\
 Caixa Postal 5008, 58051-970, Jo\~ao Pessoa, Para\'{\i}ba, Brazil}
\email{jroberto,petrov,edilberto@fisica.ufpb.br}

\begin{abstract}
Within the superfield approach, we prove the absence of UV/IR mixing in the three-dimensional noncommutative supersymmetric Maxwell-Chern-Simons theory at any loop order and demonstrate its finiteness in one, three and higher loop orders.
\end{abstract}

\maketitle

\section{Introduction}

Renormalizability is one of the more important issues in quantum field theories. In the case of the noncommutative (NC) field theories, it is accompanied by the problem of UV/IR mixing which generates infrared singularities, with no analogs in the commutative setting. Whenever stronger than logarithmic, these singularities are very dangerous as they can invalidate the usual perturbative expansion~\cite{Minw}. It is natural to expect that the supersymmetry, being known to improve the ultraviolet (UV) behavior, could also solve the problem posed by the UV/IR mixing. That this indeed is the case was explicitly shown first for gauge theories in~\cite{matusis} and for scalar theories in~\cite{WZ0,WZ2,WZ}). Several supersymmetric noncommutative models were studied since then and shown to be free of dangerous UV/IR singularities, both in three and four dimensional spacetimes~\cite{zanon1,sig,sig1,ourqed,scpn,trava}. Supersymmetric noncommutative gauge theories were particularly much studied, and several interesting properties were found, for example, nontrivial constraints arising from gauge symmetry both at the classical~\cite{mat,chaichian} and at the quantum level~\cite{zanon2,trek,oursym3,oursym4}. Moreover,~\cite{JJ,JJ2} it was argued that maximally supersymmetric noncommutative gauge theories in four spacetime dimensions were ultraviolet finite, similarly to their commutative counterparts; some explicit calculations, including considerations on the UV/IR singularities in such maximally supersymmetric theories can be found in~\cite{trava,oursym4}. 

In three spacetime dimensions, supersymmetric gauge theories are super-renormalizable and, therefore, nice candidates to be finite quantum field theories. In the noncommutative case, this possibility was partially explored in~\cite{ourqed3,oursym3}, both for Abelian and non-Abelian gauge groups. It was shown that the noncommutative supersymmetric QED and Yang-Mills theories (coupled to matter or not) were indeed finite at one loop, while power counting indicates the absence of superficial divergences at three loops or more. In this work, we perform a similar analysis for the three-dimensional supersymmetric noncommutative Maxwell-Chern-Simons (MCS) theory and explicitly show its finiteness in one, three and higher loop orders. This model is, therefore, another example of a nontrivial (interacting) quantum field theory which is actually ultraviolet finite in perturbation theory.

\section{Classical action of the theory}
The action of the Maxwell-Chern-Simons theory is (following the notations of~\cite{SGRS}),
\begin{eqnarray}
S&=&\frac{1}{2g^2}\int d^5 z \Big[W^\alpha *W_\alpha+m\Big( A^{\alpha}*W_{\alpha}+\nonumber\\&+&\frac{i}{6}\{A^{\alpha},A^{\beta}\}_**D_{\beta}A_{\alpha}+\frac{1}{12}\{A^{\alpha},A^{\beta}\}_**\{A_{\alpha},A_{\beta}\}_*\Big)\Big]\,. \label{2n}
\end{eqnarray}
where
\begin{eqnarray}
\label{sstr}
W_\beta =\frac{1}{2}D^\alpha D_\beta A_\alpha -
\frac{i}{2}[A^\alpha ,D_\alpha A_\beta ]_*-
\frac{1}{6}
[A^\alpha ,\{A_\alpha ,A_\beta \}_*]_*
\end{eqnarray}
is a superfield strength constructed from the spinor superpotential $A_\alpha$. Hereafter it is implicitly assumed that all commutators and anticommutators are Moyal ones, that is, $[A,B\}_*~\equiv~A~*~B~\mp~B~*~A$, with
\begin{equation}
A(x)*B(x)\,\equiv\, A(x)\, exp\left(\frac{i}{2}\,\overleftarrow{\frac{\partial}{\partial
x^{\mu}}}\,\Theta^{\mu\nu}\,\overrightarrow{\frac{\partial}{\partial
x^{\nu}}}\right)\,B(x)
\end{equation}

\noindent 
being the Moyal-Groenewald $*$-product. Here, $\Theta^{\mu\nu}$ is the antisymmetric real constant
matrix characterizing the noncommutativity of the underlying space-time. It is customary to impose that $\Theta^{0i}=0$ to avoid causality and unitarity issues~\cite{Gomis}, even if these problems can be solved by alternative formulations of the noncommutativity~\cite{Bahns,Liao,Balachandran}. In our case, however, we will show that the choice $\Theta^{0i}=0$ will also play an essential role in ensuring the absence of infrared UV/IR singularities. 

The action in Eq.~(\ref{2n}) is invariant under the infinitesimal gauge transformations
\begin{eqnarray}
\label{gt}
\delta A_{\alpha}=D_{\alpha}K-i[A_{\alpha},K]\,.
\end{eqnarray}

\noindent
After gauge fixing, the total action of the noncommutative supersymmetric MCS theory reads
\begin{eqnarray}
\label{st}
S_{total}=S+S_{GF}+S_{FP}\,,
\end{eqnarray}

\noindent
where $S_{GF}$ is the gauge fixing term,
\begin{eqnarray}
S_{GF}=-\frac{1}{4\xi g^2}\int d^5 z (D^\alpha A_\alpha )
D^2(D^\beta A_\beta )\,,
\end{eqnarray}
and $S_{FP}$ the corresponding action for the Faddeev-Popov ghosts,
\begin{eqnarray}
\label{sfp}
S_{FP}=\frac{1}{2g^2}\int d^5 z (c'D^\alpha D_\alpha c+ic'*
D^\alpha [A_\alpha ,c])\,.
\end{eqnarray}

From Eq.~(\ref{st}), one obtains the free spinor superpotential propagator,
\begin{eqnarray}
<A^\alpha (-p,\theta_1)A^\beta (p,\theta_2)>=\frac{1}{i}g^2\left[-\frac{(D^2-m)D^{\beta}D^{\alpha}}{2p^2(p^2+m^2)}+\xi\frac{D^2D^{\alpha}D^{\beta}}{2p^4}
\right]\delta_{12}\,,
\end{eqnarray}
and the propagator for the ghosts fields,
\begin{eqnarray}
\label{pr2}
<c'(-k,\theta_1)c(k,\theta_2)>=ig^2\frac{D^2}{k^2}\delta_{12}\,,
\end{eqnarray}
where, $\delta_{12}=\delta^2(\theta_1-\theta_2)$. Also, whenever not otherwise indicated, it must be understood that the supercovariant derivatives act on the Grassmann variable $\theta_1$). Finally, also from Eq.~(\ref{st}) one can extract the interaction part of the classical action in the pure gauge sector,
\begin{eqnarray}
\label{sint}
S_{int}&=&\frac{1}{g^2}\int d^5 z\Big[-\frac{i}{4}D^{\gamma}D^\alpha
A_{\gamma}*
[A^\beta ,D_\beta A_\alpha ]-\frac{1}{12}D^{\gamma}D^\alpha A_{\gamma}*
[A^\beta ,\{A_\beta ,A_\alpha \}]-\nonumber\\&-&
\frac{1}{8}[A^{\gamma},D_{\gamma}A^\alpha ]*[A^\beta ,D_\beta A_\alpha
]+\frac{i}{12}
[A^{\gamma},D_{\gamma}A^\alpha ]*[A^\beta ,\{A_\beta ,A_\alpha \}]+
\nonumber\\&+&\frac{1}{72}
[A^{\gamma},\{A_{\gamma},A^\alpha \}]*[A^\beta ,\{A_\beta ,A_\alpha \}]+\nonumber\\&+&
m\Big(A^{\alpha}*\Big(\frac{i}{2}[A^\alpha ,D_\alpha A_\beta ]-
\frac{1}{6}[A^\alpha ,\{A_\alpha ,A_\beta \}]\Big)+\frac{i}{6}\{A^{\alpha},A^{\beta}\}*D_{\beta}A_{\alpha}+\nonumber\\&+&\frac{1}{12}\{A^{\alpha},A^{\beta}\}*\{A_{\alpha},A_{\beta}\}
\Big)\Big]\,,
\end{eqnarray}

\noindent
from which the interacting vertices for the perturbative calculations can be read directly.

\section{Cancellation of the one-loop divergences in the pure gauge sector}

Now, let us study the one-loop divergences in the theory. Similarly to~\cite{ourqed3}, the superficial degree of the divergence can be found to be
\begin{eqnarray}
\label{o}
\omega=2-\frac{1}{2} V_c-2V_0-\frac{3}{2}V_1-V_2-\frac{1}{2} V_3-\frac{1}{2} N_D\,,
\end{eqnarray}
where $V_i$ is the number of purely gauge vertices involving $i$ supercovariant derivatives, $V_c$ is the number of gauge-ghost vertices, and $N_D$ is the number of the spinor derivatives acting on the external fields. This power counting relationship characterizes the Maxwell-Chern-Simons theory as an UV super-renormalizable theory. It is easy to realize that linear divergences may come only from
graphs with  $V_3=2$, or $V_2=1$,  or $V_c=2$. Since the Chern-Simons sector involves only vertices with at the most one spinor derivative, we conclude that these vertices do not contribute to possible linear divergences.  

\begin{figure}[ht]
\includegraphics{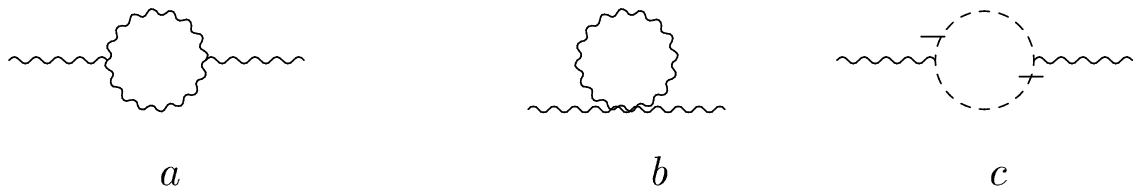}
\caption{Superficially linearly divergent diagrams contributing to the
two-point function of the gauge field in the purely gauge sector.}
\label{f.1}
\end{figure}

The potentially linearly divergent graphs are depicted in  Fig.~\ref{f.1}, they contribute to the two-point functions of $A^{\alpha}$ field. In these graphs, a cut in a ghost line corresponds to the factor $D_{\alpha}$ acting on the ghost propagator. A trigonometric factor $e^{ik\wedge
l}-e^{il\wedge k}=2i\sin(k\wedge l)$, where $k\wedge l\equiv k^{\mu}l^{\nu}\Theta_{\mu\nu}$, originates from each commutator. By denoting the contributions of the graphs in Fig.~\ref{f.1} by $I_{1a}$, $I_{1b}$,  and $I_{1c}$, respectively, we have (the use of dimensional reduction is implicitly assumed hereafter)
\begin{eqnarray}
I_{1a}&=&-\frac{1}{32}\int \frac{d^3p}{(2\pi)^3}d^2 \theta_1 d^2\theta_2
\int\frac{d^3k}{(2\pi)^3}\sin^2(k\wedge p)
A^\beta (-p,\theta_1)A^{\beta^\prime}(p,\theta_2)\times
\nonumber\\&\times&
\left\{D^{\gamma}D^{\alpha}
\left[-\frac{(D^2-m)D_{\gamma^{\prime}}D_{\gamma}}{k^2(k^2+m^2)}+\xi\frac{D^2D_{\gamma}D_{\gamma^{\prime}}}{k^4}
\right]D^{\alpha^{\prime}}D^{\gamma^{\prime}}
\delta_{12}\right.\times\nonumber\\&\times&\left.D_{\beta}
\left[-\frac{(D^2-m)D_{\alpha^{\prime}}D_{\alpha}}{(p-k)^2((p-k)^2+m^2)}+\xi\frac{D^2D_{\alpha}D_{\alpha^{\prime}}}{(p-k)^4}
\right]D_{\beta^{\prime}}
\delta_{12}\,-\right.
\nonumber\\&-&\left.
D^{\gamma}D^{\alpha}
\left[-\frac{(D^2-m)D_{\gamma^{\prime}}D_{\gamma}}{k^2(k^2+m^2)}+\xi\frac{D^2D_{\gamma}D_{\gamma^{\prime}}}{k^4}
\right]D_{\beta^{\prime}}
\delta_{12}\right.\times\nonumber\\&\times&\left.D_{\beta}
\left[-\frac{(D^2-m)D_{\alpha^{\prime}}D_{\alpha}}{(p-k)^2((p-k)^2+m^2)}+\xi\frac{D^2D_{\alpha}D_{\alpha^{\prime}}}{(p-k)^4}
\right]D^{\alpha^{\prime}}D^{\gamma^{\prime}}
\delta_{12}
\right\}\,+\,\cdots\,,
\label{mlett:a1}
\end{eqnarray}
\begin{eqnarray}
I_{1b}&=&-\frac{1}{3}\int \frac{d^3p}{(2\pi)^3}d^2 \theta_1
\int\frac{d^3k}{(2\pi)^3}\sin^2(k\wedge p)
\nonumber\\&\times&\Big[A^{\beta}(-p,\theta_1)A_{\alpha}(p,\theta_1)D^{\gamma}D^{\alpha}[-\frac{(D^2-m)D_{\beta}D_{\gamma}}{2k^2(k^2+m^2)}+\xi\frac{D^2D_{\gamma}D_{\beta}}{2k^4}]\delta_{12}|_{\theta_1=\theta_2}+\nonumber\\&+&
A^{\beta}(-p,\theta_1)A_{\beta}(p,\theta_1)D^{\gamma}D^{\alpha}[-\frac{(D^2-m)D_{\alpha}D_{\gamma}}{2k^2(k^2+m^2)}+\xi\frac{D^2D_{\gamma}D_{\alpha}}{2k^4}]\delta_{12}|_{\theta_1=\theta_2}
\Big]
\nonumber\\
&+&\frac{1}{4}\int \frac{d^3p}{(2\pi)^3}d^2 \theta A^{\gamma}(-p,\theta)A^{\beta}(p,\theta)\int\frac{d^3k}{(2\pi)^3}\sin^2(k\wedge p)
\times\nonumber\\&\times&
D_{\gamma}
\left[-\frac{(D^2-m)D_{\alpha}D^{\alpha}}{k^2(k^2+m^2)}+\xi\frac{D^2D^{\alpha}D_{\alpha}}{k^4}
\right]
D_{\beta}\delta_{12}|_{\theta_1=\theta_2},\label{mlett:b1}
\end{eqnarray}
\begin{eqnarray}
I_{1c}&=&\frac{1}{2}\int \frac{d^3p}{(2\pi)^3}d^2 \theta_1 d^2\theta_2
\int\frac{d^3k}{(2\pi)^3}\frac{\sin^2(k\wedge p)}{k^2(k+p)^2}
A_\alpha (-p,\theta_1)A_\beta (p,\theta_2)\nonumber\\
&\times&
D^\alpha _1D^2\delta_{12}D^2D^\beta _2\delta_{12}\,.
\label{mlett:c1}
\end{eqnarray}
In the expressions for the $I_1$'s the terms where covariant derivatives act on external fields were omitted
because they do not produce linear divergences and UV/IR mixing (as we shall shortly verify, such terms give only finite contributions). In the formulae above they are indicated by the ellipsis. After some D-algebra transformations we arrive at
\label{contrib1}
\begin{eqnarray}
I_{1a}&=&\frac{1}{2}\xi\int \frac{d^3p}{(2\pi)^3}d^2 \theta_1
\int\frac{d^3k}{(2\pi)^3}\frac{\sin^2(k\wedge p)}{k^2}
A^\beta (-p,\theta_1)A_\beta (p,\theta_1)
\,+\,\cdots\,,\label{mlett:acontrib1}\\
I_{1b}&=&\frac{1}{2}(1-\xi)\int \frac{d^3p}{(2\pi)^3}d^2 \theta_1
\int\frac{d^3k}{(2\pi)^3}\frac{\sin^2(k\wedge p)}{k^2}
A^\beta (-p,\theta_1)A_\beta (p,\theta_1)\,+\,\cdots\,,\label{mlett:bcontrib1}\\
I_{1c}&=&-\frac{1}{2}\int \frac{d^3p}{(2\pi)^3}d^2 \theta_1
\int\frac{d^3k}{(2\pi)^3}\frac{\sin^2(k\wedge p)}{k^2}
A^\beta (-p,\theta_1)A_\beta (p,\theta_1)\,+\,\cdots\,.\label{mlett:ccontrib1}
\end{eqnarray}
Hence, the total one-loop two-point function of the gauge superfield, given by $I_1=I_{1a}+I_{1b}+I_{1c}$, is free from both UV and UV/IR infrared linear singularities.  

It is also easy to show that the logarithmically UV divergent parts coming from the planar parts  of $I_{1a}$, $I_{1b}$ and $I_{1c}$, which involve derivatives of the gauge fields, as well as the logarithmically UV divergent parts generated by other one-loop graphs, turn out to be proportional to the integral
\begin{eqnarray}
\label{intl}
\int\frac{d^3k}{(2\pi)^3}\frac{k_{\alpha\beta}}{k^2(k+p)^2}
\end{eqnarray}
and are therefore finite by symmetric integration. Thus, the UV logarithmic singularities are also absent, i.e., the two-point function of $A^{\alpha}$ field is UV {\em finite} in the one-loop approximation. We note also the absence of the one-loop nonplanar logarithmic IR singularities. Indeed, the typical logarithmically IR singular contribution which could arise is proportional to a linear combination of integrals of the form
\begin{eqnarray}
\label{intl2}
\int\frac{d^3k}{(2\pi)^3}\frac{k_{\alpha\beta}\sin(2k\wedge p)}{k^4}\,=\,
-\frac{i}{4\pi}\frac{\tilde{p}_{\alpha\beta}}{\sqrt{\tilde{p}^2}}\,.
\end{eqnarray}
Here, $\tilde{p}_{\alpha\beta}=\Theta_{\mu\nu}p^\nu(\sigma^\mu)_{\alpha\beta}$. Remembering that $\Theta_{0i}=0$, one realizes that this last expression does not produce logarithmic divergences. The conclusion is, therefore, that all one-loop quantum corrections are actually UV finite, and no UV/IR mixing appears.

We already mentioned that linear divergences are possible only for $V_2=1$, or $V_3=2$, or
$V_c=2$. It is easy to see that two-loop graphs satisfying these conditions are just vacuum ones whereas higher-loop graphs cannot satisfy these conditions at all. Therefore, there are no linear UV and UV/IR infrared divergences beyond one-loop and, as consequence, {\em the Green functions are free of nonintegrable infrared divergences at any loop order}.

One can verify that Eq. (\ref{o}) implies in the absence of any divergences at three- and higher-loop orders, in agreement with the super-renormalizability of the theory. This concludes our analysis of the ${\cal N}=1$ supersymmetry. To complete the study of renormalization in the theory it remains to investigate the situation on the two-loop order.

\section{Cancellation of the one-loop divergences in the matter sector}

We next study the interaction of the spinor gauge field with matter. To this end we add to Eq.~(\ref{st}) the action of the $N$ scalar matter fields  $\phi_a$, with $a=1,\ldots,N$.
\begin{eqnarray}
\label{acmat}
S_m&=&\int d^5 z
\Big[-\bar{\phi}_a(D^2-m)\phi_a+i\frac{1}{2} ([\bar{\phi}_a,A^\alpha ]*D_\alpha \phi_a-
D_\alpha \bar{\phi}_a*[A^\alpha ,\phi_a])+\nonumber\\&+&
\frac{1}{2} [\bar{\phi}_a, A^\alpha] *[A_\alpha,\phi_a]\Big]\,,
\end{eqnarray}

\noindent
The free propagator of the scalar fields is
\begin{eqnarray}
<\bar{\phi}_a(-k,\theta_1)\phi_b(k,\theta_2)>=i\delta_{ab}\frac{D^2+m}{k^2+m^2}
\delta_{12}\,,
\end{eqnarray}

\noindent
and the superficial degree of divergence when matter fields are present can be shown to be equal to
\begin{eqnarray}
\label{om}
\omega=2-\frac{1}{2} V_c-2V_0-\frac{3}{2}V_1-V_2-\frac{1}{2} V_3
-\frac{1}{2} E_{\phi}-\frac{1}{2}
V^1_{\phi}-\frac{1}{2} N_D-V^0_{\phi}\,,
\end{eqnarray}
where, as before, $V_i$ is the number of pure gauge vertices with $i$
spinor derivatives, $E_{\phi}$ is the number of external scalar lines,
$N_D$ is the number of spinor derivatives associated to external lines, $V^1_{\phi}$ is the
number of triple vertices involving matter, and $V^0_{\phi}$ is the number of quartic vertices involving matter.

It is straightforward to show that the graphs with non-zero number of the external matter legs, possessing $E_{\phi}\leq 2$ together with $V^1_{\phi}>0$ or $V^0_{\phi}>0$, in the worst case, are only logarithmically divergent. Applying the arguments from the previous section, one can convince oneself that in this case the one-loop graphs are finite.

\begin{figure}[ht]
\includegraphics{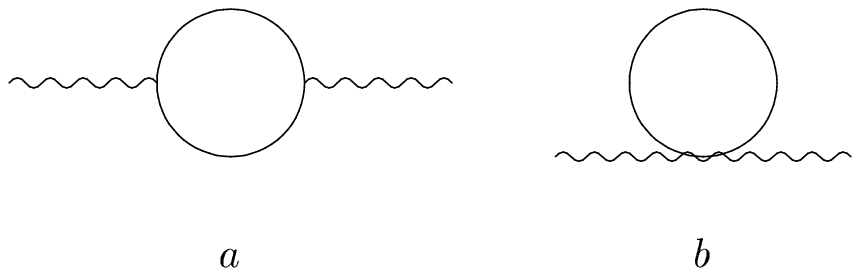}
\caption{Superficially linearly divergent diagrams contributions to the
two-point function of the gauge field  coming from the matter sector.}
\label{f.2}
\end{figure}

It remains to study the graphs with zero number of external matter legs. The leading UV divergence for them 
is $\omega = 3/2$ (one external $A_{\alpha}$ leg), corresponding to a tadpole graph which vanishes identically. What comes next are graphs with two external $A_{\alpha}$ legs which are superficially UV
linearly divergent. They are depicted in Fig.~\ref{f.2}, and their contribution was earlier found in \cite{scpn,ourqed3}, so here we merely quote the result,
\begin{eqnarray}
\label{stot}
I_4 &=& 2N \int \frac{d^3p}{(2\pi)^3} d^2\theta \int \frac{d^3k}{(2\pi)^3}
\frac{\sin^2(k\wedge p)}{(k^2+m^2)\left[(k+p)^2+m^2\right]}
\nonumber\\&\times&
(k_{\gamma\beta}-mC_{\gamma\beta})\Big[(D^2A^{\gamma}(-p,\theta)) A^{\beta}(p,\theta)
+\frac{1}{2} D^{\gamma}D^{\alpha}A_{\alpha}(-p,\theta) A^{\beta}(p,\theta)
\Big]\,.
\end{eqnarray}

As we see, the dangerous linear divergences in fact do not appear, whereas, by using the arguments of the previous section,  the logarithmic divergences are also absent. Therefore, the two-point function of $A^{\alpha}$ field turns out to be free of UV/IR infrared singularities. This two-point function can be used for deriving the effective propagators in the ${1}/{N}$ expansion~\cite{scpn}. All other one-loop graphs in the matter sector are finite.

\section{Conclusions}

Let us briefly describe the main results of the paper. We have shown that the three-dimensional noncommutative supersymmetric Maxwell-Chern-Simons theory is one-loop UV and UV/IR infrared finite both without and with matter. Similarly to~\cite{oursym3}, one can prove that the same result holds in the non-Abelian case for the gauge group  $U(N)$. A natural development of this work consists in the investigation of the
two-loop corrections to the effective action, as we pointed out earlier. It would also be interesting to study the $1/N$ expansion for the model involving many scalar fields and to analyze of spontaneous symmetry breaking and the Higgs mechanism in this class of models.

{\bf Acknowledgments.}
This work was partially supported by Funda\c c\~ao de Amparo \`a Pesquisa do Estado de S\~ao Paulo (FAPESP) and Conselho Nacional de Desenvolvimento Cient\'\i fico e Tecnol\'ogico (CNPq). The work by A. Yu. P. has been supported by CNPq-FAPESQ DCR program, CNPq project No. 350400/2005-9, while A. F. F. acknowledges the support from FAPESP, project No. 04/13314-4.

\newpage

\begin{thebibliography}{10}

\bibitem{Minw} S. Minwalla, M. van Raamsdonk, N. Seiberg, JHEP {\bf 02}, 020 (2000).

\bibitem{matusis} A. Matusis, L. Susskind, N. Toumbas, JHEP {\bf 12}, 002 (2000).

\bibitem{WZ0} H. O. Girotti, M. Gomes, V. O. Rivelles, A. J. da Silva, Nucl. Phys. {\bf B587}, 299 (2000).

\bibitem{WZ2} A. A. Bichl, J. M. Grimstrup, H. Grosse, L. Popp, M. Schweda, R. Wulkenhaar, JHEP {\bf 10}, 046 (2000). 

\bibitem{WZ} I. L. Buchbinder, M. Gomes, A. Yu. Petrov, V. O. Rivelles, Phys. Lett. {\bf B517}, 191 (2001).

\bibitem{zanon1} D. Zanon, Phys. Lett. {\bf B502}, 265 (2001).

\bibitem{sig}  H. O. Girotti, M. Gomes, V. O. Rivelles, A. J. da Silva,  Int. J. Mod. Phys. {\bf A17}, 1503 (2001).

\bibitem{sig1} H. O. Girotti, M. Gomes, A. Yu. Petrov, V. O. Rivelles, A. J. da Silva. Phys. Lett. {\bf B521}, 119 (2001).

\bibitem{ourqed}  A. F. Ferrari, H. O. Girotti, M. Gomes, A. A. Ribeiro, A. Yu. Petrov, V. O. Rivelles, A.J. da Silva,
Phys. Rev. {\bf D69} 025008 (2004).

\bibitem{scpn} E. A. Asano, H. O. Girotti, M. Gomes, A. Yu. Petrov, A. G. Rodrigues, A. J. da Silva, 
Phys. Rev. {\bf D69}, 105012 (2004).

\bibitem{trava} V. V. Khoze, G. Travaglini, JHEP {\bf 01}, 026 (2001).

\bibitem{mat} K. Matsubara, Phys. Lett. {\bf B482}, 417 (2000).

\bibitem{chaichian} M. Chaichian, P. Presnajder, M. M. Sheikh-Jabbari, A. Tureanu, Phys. Lett. {\bf B526}, 132 (2002).

\bibitem{zanon2} M. Pernici, A. Santambrogio, D. Zanon, Phys.Lett. {\bf B504}, 131 (2001).

\bibitem{trek} H. Liu, J. Michelson, Nucl. Phys. {\bf B614}, 279 (2001).

\bibitem{oursym3} A. F. Ferrari, H. O. Girotti, M. Gomes, A. A. Ribeiro, A. Yu. Petrov, A.J. da Silva,
Phys. Lett. {\bf B601}, 88 (2004).

\bibitem{oursym4}  A. F. Ferrari, H. O. Girotti, M. Gomes, A. Yu. Petrov, A. A. Ribeiro, V. O. Rivelles, A. J. da Silva,
Phys. Rev. {\bf D70}, 085012 (2004).

\bibitem{JJ} I. Jack, D. R. T. Jones, Phys. Lett. {\bf B514}, 401 (2001).

\bibitem{JJ2} I. Jack, D. R. T. Jones, New. J. Phys. {\bf 3}, 19 (2001).

\bibitem{ourqed3} A. F. Ferrari, H. O. Girotti, M. Gomes, A. Yu. Petrov,
A. A. Ribeiro, A. J. da Silva,  Phys. Lett. {\bf B577}, 83 (2003).

\bibitem{SGRS} S. J. Gates, M. T. Grisaru, M. Rocek, W. Siegel. Superspace or One Thousand and One Lessons in Supersymmetry. Benjamin/Cummings, 1983.
 
\bibitem{Gomis} J. Gomis and T. Mehen, Nucl. Phys. {\bf B591}, 265 (2000).

\bibitem{Bahns} D.~Bahns, S.~Doplicher, K.~Fredenhagen, G.~Piacitelli, Phys.Lett. {\bf B533}, 178 (2002).

\bibitem{Liao} Y.~Liao, K.~Sibold, Eur. Phys. J. {\bf C25}, 469 (2002).

\bibitem{Balachandran}
A.~P.~Balachandran, T.~R.~Govindarajan, C.~Molina, P.~Teotonio-Sobrinho, JHEP {\bf 10}, 072 (2004).

\end{thebibliography}
\end{document}